\documentclass[aps,twocolumn,superscriptaddress,PRB]{revtex4-1}

\usepackage[utf8]{inputenc}

\usepackage{natbib}
\usepackage{graphicx}
\usepackage{dcolumn}
\usepackage{bm}
\usepackage{epsfig}
\usepackage{textcomp}
\usepackage{gensymb}

\begin{document}

\title{Surface Structure Determination of Black Phosphorus Using Photoelectron Diffraction}

\author{Luis Henrique de Lima}
\affiliation{Instituto de F\'{i}sica Gleb Wataghin, Universidade Estadual de Campinas, Campinas 13083-859, SP, Brazil}

\author{Lucas Barreto}
\affiliation{Centro de Ciências Naturais e Humanas, Universidade Federal do ABC, Santo André 09210-580, SP, Brazil}
\affiliation{Instituto de F\'{i}sica Gleb Wataghin, Universidade Estadual de Campinas, Campinas 13083-859, SP, Brazil}

\author{Richard Landers}
\affiliation{Instituto de F\'{i}sica Gleb Wataghin, Universidade Estadual de Campinas, Campinas 13083-859, SP, Brazil}

\author{Abner de Siervo}
\affiliation{Instituto de F\'{i}sica Gleb Wataghin, Universidade Estadual de Campinas, Campinas 13083-859, SP, Brazil}

\begin{abstract}
Atomic structure of single-crystalline black phosphorus was studied by high resolution synchrotron-based photoelectron diffraction (XPD). The results show that the topmost phosphorene layer in the black phosphorus is slightly displaced compared to the bulk structure and presents a small contraction in the direction perpendicular to the surface. Furthermore, the XPD results show the presence of a small buckling among the surface atoms, in agreement with previously reported scanning tunneling microscopy results. The contraction of the surface layer added to the presence of the buckling indicates an uniformity in the size of the $sp^3$ bonds between P atoms at the surface.
\end{abstract}

\maketitle

\begin{figure}[!h]
     \centering
     \includegraphics[scale=0.28]{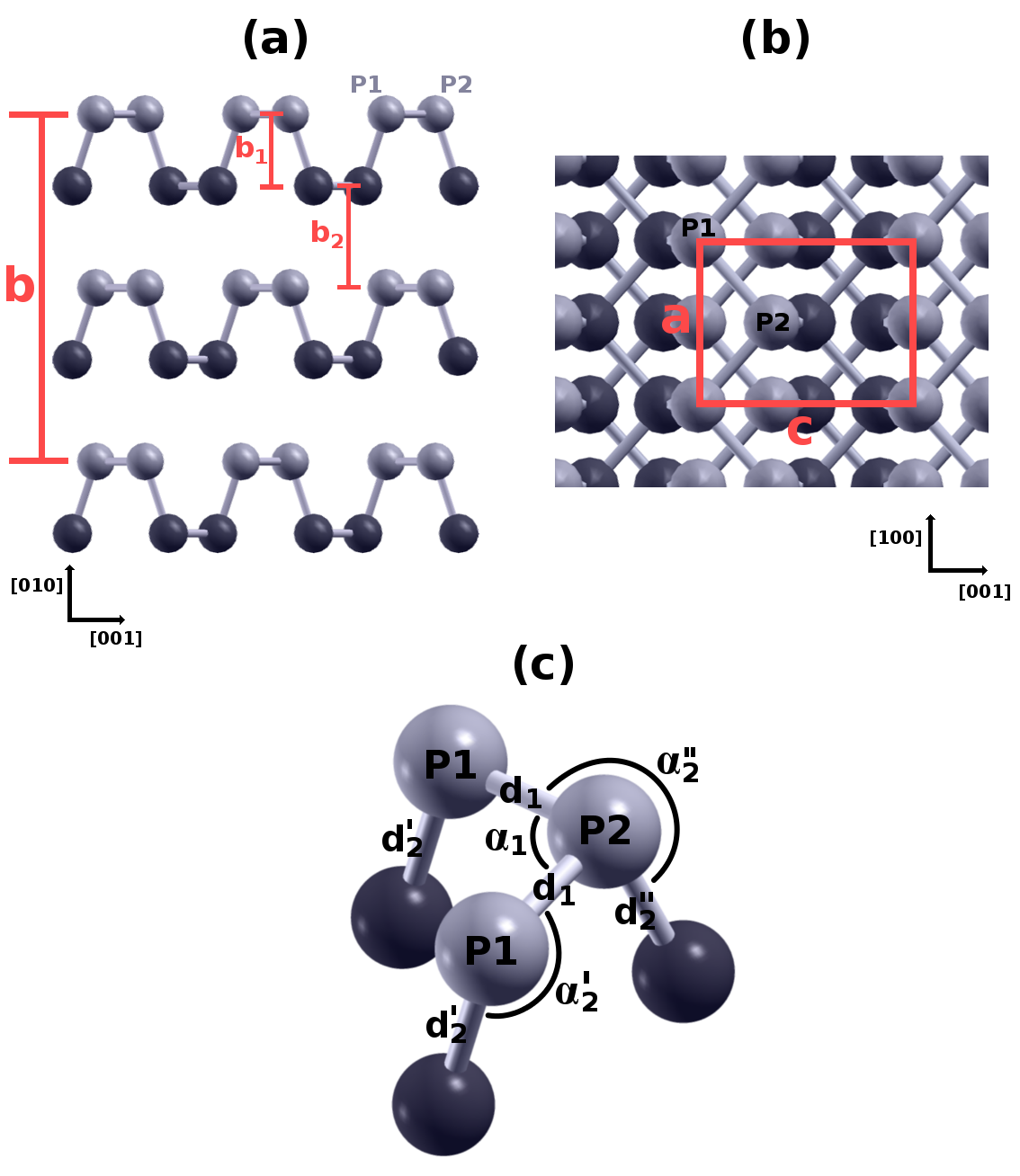}
     \caption{Schematic illustration of the BP atomic structure: (a) Side view ({\it bc} plane). It is shown three bilayers (or three phosphorene layers). The P atoms are shown in different gray scales for clarity. (b) Top view ({\it ac} plane). It is shown two bilayers and the in-plane unit cell. (c) Nearest neighbour distances and bond angles for the surface phosphorene layer. There are two distinct d$_2$ and $\alpha_2$ due to the buckling, as explained in the text.}
     \label{fig1}
\end{figure}

Since the experimental advent of graphene \cite{Novoselov2004}, other 2D materials have received enormous attention due to their great potential in nanoscale devices \cite{Xu2013}. The 2D layered materials are characterized by atoms making strong covalent in-plane bonds, but the stacking of these atomic layers resulting from relatively weak interactions of van-der-Waals type. Besides graphene, other examples of 2D materials are hexagonal boron nitride ({\it h}-BN) and transition metal dichalcogenides (TMDs), for example, MoS$_2$, MoSe$_2$, WSe$_2$, WS$_2$, among others \cite{Xu2013}. Another interesting possibility is the design of heterostructures from the stacking of different monolayers of 2D materials, with these new materials presenting distinct properties \cite{Geim2013}. Recently, orthorhombic black phosphorus (BP), the most stable phosphorus allotrope, has emerged as a ``new'' promising material for applications in nanoelectronics and nanophotonics \cite{Ling2015}. The BP is formed by a stack of phosphorus layers arranged in a honeycomb structure \cite{Brown1965,Morita1986} known as phosphorene. As usual in 2D materials, the phosphorene layers are held together by a weak interaction, which allows the mechanical exfoliation procedure similar to that applied to graphene \cite{Favron2015}. However, unlike graphene, where the carbon monolayer is strictly flat, the phosphorene has a strongly puckered structure, where each phosphorene layer can be seen as a bilayer of P atoms, as shown in Fig. \ref{fig1}a and \ref{fig1}b. Within the phosphorene layer, each atom is covalently bonded to three neighbours (sp$^3$ hybridization), with two bonds connecting the nearest P atoms in the same plane, and the third bond conecting P atoms between the top and bottom of the phosphorene layer, as shown in Fig. \ref{fig1}c. Another important difference between BP and graphene is the presence of a direct band gap, which is theoretically expected to vary with the number of phosphorene layers from $\sim$0.3 eV for bulk BP to $\sim$2 eV for the single layer \cite{Tran2014}. The theoretical position of the band gap is a subject of some controversy in the literature \cite{Asahina1982,Goodman1983,Li2014,Du2010}. From the experimental point of view, angle-resolved photoemission (ARPES) results show that the band gap is located at the $Z$ point of the Brillouin Zone \cite{Li2014,Takahashi1984,Takahashi1986,Han2014}.

Another interesting aspect involves the band gap engineering. Several theoretical results for the bulk, few layers, phosphorene and nanoribbons show that it is possible to tune the energy and the position of the band gap by strain and application of an electric field \cite{Rodin2014,Li2014_A,Peng2014,Dai2014,Guo2014,Liu2015,Wu2015,Fei2015}. This controlled modification of the electronic structure plays a fundamental role in a possible future application of BP \cite{Ling2015}, especially using an electric field, which is more feasible in gated devices. Experimentally, the effect of applying an electric field was carried out by doping the material with alkali metals \cite{Kim2015}, similarly to what was done for graphene \cite{Ohta2006}. Band gap modulation together with an electronic band inversion for different concentrations of K atoms were observed \cite{Kim2015}, with the BP evolving from a moderate gap semiconductor to a Dirac semimetal, and a linear dispersion observed in the [001] in-plane direction (see Fig. \ref{fig1}b) \cite{Kim2015}, as predicted theoretically \cite{Fei2015}.

From the electronic point of view, the cited examples demonstrate that the BP has been extensively characterized and studied, both theoretically and experimentally. On the other hand, from the atomic structure point of view and particularly for the surface, the number of studies is limited. The pioneering studies focused on determining the bulk atomic structure, for example, by X-ray diffraction \cite{Brown1965} and neutron powder diffraction \cite{Cartz1979}. Recent results using scanning transmission electron microscopy \cite{Wu2015_A} were also obtained for the bulk, with good agreement with the experimental results of X-ray diffraction. For the BP surface, a combined study using scanning tunneling microscopy (STM) and density functional theory (DFT) calculations \cite{Zhang2009} show that the surface atoms occupy almost the same position of the atoms in the bulk, except for a small perpendicular relaxation of the P1 and P2 surface atoms (see Fig. \ref{fig1}).

Although a large number of studies with theoretical predictions for phosphorene have been published recently \cite{Rodin2014,Peng2014,Liu2015,Wu2015,Aierken2015}, the experimental results reported use in general bulk BP samples \cite{Zhang2009,Han2014,Kim2015,Edmonds2015} or few layers obtained by exfoliation \cite{Xia2014,Wu2015_A,Favron2015}. The efficient production of a phosphorene single layer with its atomic structure and orientation characterized is still a technological challenge \cite{Ling2015}. Recently published works demonstrate that the electrical and thermal properties of the single layer are spatially in-plane anisotropic \cite{Fei2015,Kim2015,Aierken2015}, and therefore it is fundamental to know the crystallographic orientation of the BP when inserted into a device in order to take advantage of these properties \cite{Xia2014}.

Despite the difficulty and challenge in the characterization of the atomic structure of an isolated single layer, the surface of a single crystal can be a good approximation, since the surface is a natural break in the perpendicular periodicity. Notwithstanding some questions remain, for example, how different is the distance indicated as $b_1$ in Fig. \ref{fig1}a compared to the bulk? Another question that arises concerns the small buckling observed in STM images \cite{Zhang2009}. Does this buckling observed in STM exist or it is just an electronic artifact of the imaging? In order to answer these questions, a structural determination of the BP surface using high resolution synchrotron-based photoelectron diffraction (XPD) is extremely appropriate. We shall stress the importance of such experimental result, since most of the bulk measurements were done under atmospheric conditions, and STM is not very precise in determining interlayer distances.  Moreover, most of the electronic results already predicted are based on models which assume the bulk structure. Surface relaxation or reconstruction, for instance showing a buckling behavior for different P atoms at the surface, might have an impact in the calculated results for the electronic structure. Therefore an experimental input from a very precise surface structure determination is needed for a complete understanding of the material and the results are presented in the following.  

\begin{figure}[!h]
     \centering
     \includegraphics[scale=0.3]{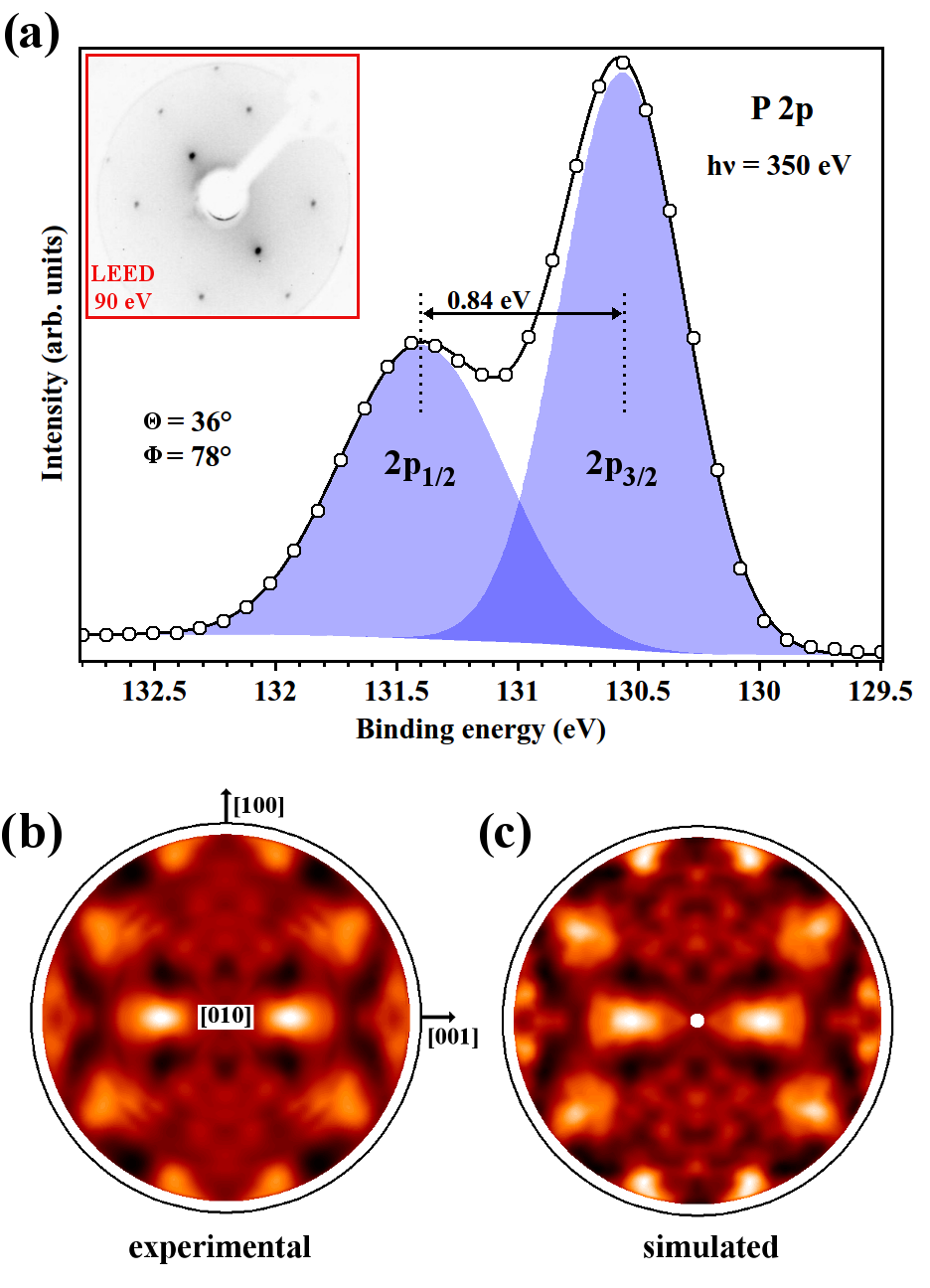}
     \caption{(a) P 2p core-level spectrum recorded at photon energy of h$\nu$ = 350 eV and at polar and azimuth angles of $\Theta = 36^{\circ}$ and $\Phi = 78^{\circ}$, respectively. The continuous black line represents the fitting envelope consisting of the 2p$_{1/2}$ and 2p$_{3/2}$ components. The open dots are the experimental data. The inset shows a LEED pattern measured with 90 eV electron energy. (b) Experimental photoelectron diffraction pattern. The main crystallographic directions are shown. (c) Simulated photoelectron diffraction pattern. The patterns are orthographic projections.}
     \label{fig2}
\end{figure}

The angle-scanned XPD experiments were carried out at the PGM beamline of the Brazilian Synchrotron Light Laboratory (LNLS) \cite{Cezar2013}. A commercial BP single crystal (HQ Graphene) was cleaved in high vacuum, \mbox{P = $3\times10^{-7} $ mbar}, using a scotch tape and transferred immediately (few seconds) to the analysis chamber with the pressure maintained below \mbox{$1\times10^{-10}$ mbar} during the whole photoemission measurements. It is possible to observe from the Low Energy Electron Diffraction (LEED) pattern (top left of Fig. \ref{fig2}a) the quality of the crystal, composed of a single domain with well-defined spots and the absence of diffuse background. Fig. \ref{fig2}a also shows one of the 2760 high-resolution X-ray Photoelectron Spectroscopy (XPS) spectra, which were used to construct the diffraction pattern presented in Fig. \ref{fig2}b. The P 2p core-level was probed with 350 eV photons, which results in photoelectrons with a kinetic energy of $\sim$ 220 eV. We also measured XPS spectra with higher photon energy (h$\nu$ = 650 eV) which showed no contamination of other elements, for instance, carbon or oxygen (results not shown).

The photoelectron diffraction pattern was recorded over a polar angle range of $3\degree\leq\Theta\leq69\degree$, and over a full $360\degree$ azimuthal range ($\Phi$), in steps of $3\degree$ for both angles. The polar angle $\Theta = 0\degree$ corresponds to normal emission. Fig. \ref{fig2}b shows the experimental photoelectron diffraction pattern. Each point in the pattern corresponds to the summed area of the 2p$_{1/2}$ and 2p$_{3/2}$ components in Fig. \ref{fig2}a, with the intensity modulation related to the atomic structure around the emitter atom. The diffraction pattern of Fig. \ref{fig2}c was obtained from multiple scattering calculations with the MSCD package \cite{Chen1998} allowing a maximum of 8 scattering events in a 245 atoms cluster. The simulated pattern was obtained after a relaxation process of the atomic structure using a method based on the genetic algorithm \cite{Viana2007}. The structure is determined in a search process for the best set of parameters that describes the agreement between theory and experiment through minimization of the reliability factor ({\it R-factor}), as described elsewhere \cite{deSiervo2002}. After obtaining the best structure, some structural parameters were varied around their best values in order to assess whether these changes actually produce significant changes in the value of {\it R-factor} and also to estimate the errors associated with these parameters. The error associated to the parameters was determined using the procedure reported in the literature \cite{Booth1997,Bondino2002}.  

\begin{figure}[!h]
     \centering
     \includegraphics[scale=0.25]{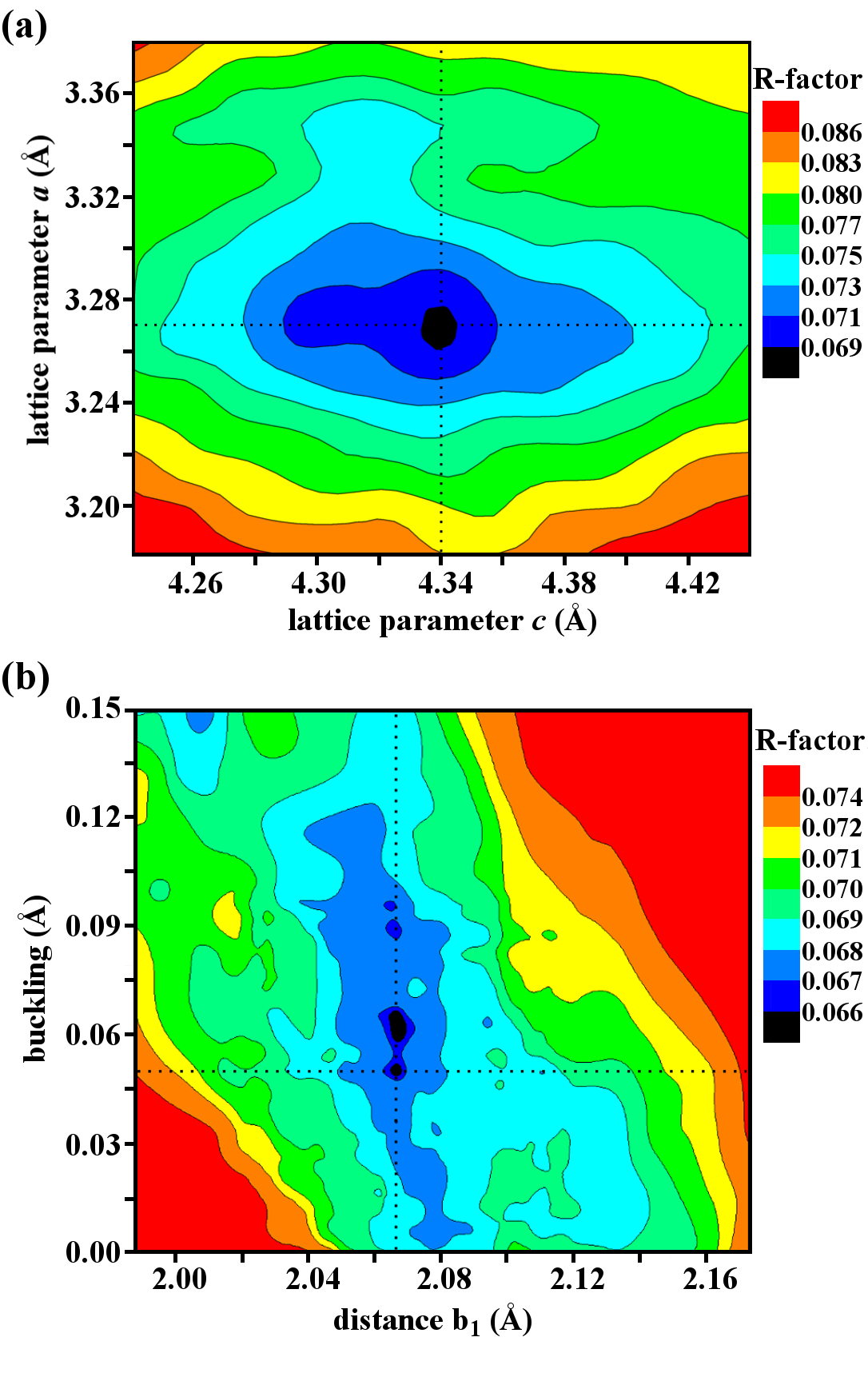}
     \caption{(a) Heat map of the {\it R-factor} as a function of the lattice parameters {\it a} and {\it c}. (b) Heat map of the {\it R-factor} as a function of the $b_1$ distance and the buckling (see Fig. \ref{fig1} and the text).}
     \label{fig3}
\end{figure}

As a starting model for the atomic structure of the BP surface, the structure expected for the bulk was used. We used the lattice parameters of the orthorhombic structure, {\it a} = 3.313 \AA, {\it b} = 10.473 \AA\ and {\it c} = 4.374 \AA, as reported by Cartz {\it et al.} \cite{Cartz1979}. The {\it R-factor} obtained with these parameters is 0.083, which is a good result and indicates that the expected structure for the surface should not differ substantially from that observed for the bulk, in agreement with the conclusions obtained by another study \cite{Zhang2009}. We will use the lattice parameters cited earlier as reference values to compare our XPD results throughout the text.

In order to refine the surface atomic structure, the structure was relaxed searching for small variations compared to the bulk, as already mentioned. Fig. \ref{fig3}a shows a heat map  of the {\it R-factor} as a function of the in-plane lattice parameters. From the map it is possible to observe a well-defined minimum, with the best values of {\it a} = 3.27(4) \AA\ and {\it c} = 4.34(6) \AA. The values obtained by Zhang {\it et al.} with STM \cite{Zhang2009} were: {\it a} = 3.33 \AA\ and {\it c} = 4.33 \AA. Therefore, the value obtained by XPD for the in-plane lattice parameter {\it c} is in excellent agreement with the results of STM and both are slightly compressed compared to the value obtained for the bulk. For the lattice parameter {\it a}, the value obtained by XPD is also slightly lower than that obtained for the bulk and differ only 1.8\% from the value obtained by STM. 

\begin{figure}[!h]
     \centering
     \includegraphics[scale=0.25]{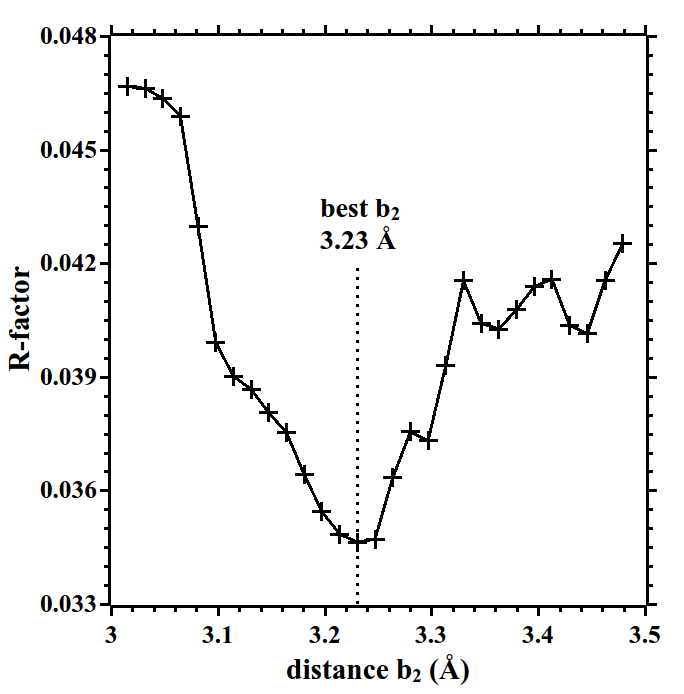}
     \caption{{\it R-factor} dependence with the $b_2$ distance (see Fig. \ref{fig1}). It is observed a small expansion compared to the expected bulk distance of 3.07 \AA\ \cite{Cartz1979}.}
     \label{fig4}
\end{figure}

In the following will be presented the values obtained for the distances parallel to the direction [010], that is, perpendicular to the surface. It is worth noting that the kinetic energy of the photoelectron is about 220 eV, which results in an inelastic mean free path of \mbox{$\sim 8$ \AA}\ \cite{nist-imfp}. This value is smaller than the lattice parameter {\it b}, which highlights the surface character of the structure determination presented here. Basically, the probed distances were only the first $b_1$ and $b_2$ distances, as shown in Fig. \ref{fig1}.

The other structural parameter analyzed is what we have called buckling. In the previously cited STM work by Zhang \cite{Zhang2009}, a slight relaxation perpendicular to the surface was observed, resulting in a height difference (buckling) of 0.02 \AA\ between P1 and P2 atoms, indicated in Fig. \ref{fig1}. The photoelectron diffraction technique has been successfully used to determine and quantify the buckling that exists in the buffer-layer obtained on the surface of SiC(0001) \cite{deLima2013,deLima2014_A} and is completely adequate to verify the presence or absence of such a buckling in the BP surface. 

Fig. \ref{fig3}b shows a heat map of the {\it R-factor} as a function of the $b_1$ distance and the buckling. The best value obtained for the $b_1$ parameter was 2.07(8) \AA, which indicates that the topmost bilayer is compressed about 5\% compared to the expected value of 2.17 \AA\ for the bulk. Also a subtle improvement in the agreement between theory and experiment was observed when the buckling was included. In this relaxation process, we included a decoupling between the two rectangular sublattices formed by the atoms P1 and P2, allowing each sublattice to move vertically and independently. The buckling was included only in the topmost atoms of the surface bilayer. The best value obtained for the buckling was 0.05(5) \AA \, which is larger than the value obtained in the STM study, but within the expected value if we take into account the experimental error. In fact, a lower sensitivity to the perpendicular distances compared to the in-plane distances was observed, as a result, there is a higher uncertainty for the perpendicular distances probed. This is clear by the presence of several local minima on the heat map of Fig. \ref{fig3}b in contrast to the behavior observed in the heat map of Fig. \ref{fig2}a. However, the variation of the {\it R-factor} as a function of the buckling along the vertical dotted line in Fig. \ref{fig3}b shows a clear trend of improvement when the buckling is taken into account. The existence of this small buckling confirms the observed contrast in STM images between the two rectangular sublattices of the surface. As argued by Zhang {\it et al.} \cite{Zhang2009}, this small variation in height is enough to observe an apparent discrimination of the electronic structure probed by STM. We observed no difference in the choice of which P1 or P2 atom is located closer/far to the inner layers, as expected. Furthermore, the definition of the distance $b_1$ is related to the sublattice closer to the inner layers.

Finally, Fig. \ref{fig4} shows the dependence of the {\it R-factor} with the $b_2$ distance. In this case, to obtain higher sensitivity, only the polar angles measured in the range $3\degree\leq\Theta\leq36\degree$ were used, since the photoelectrons emitted in these directions are those which carry more information from the deeper layers. The value obtained for $b_2$ was 3.23 \AA, which is \mbox{$\sim$ 5\%} greater than the expected bulk value (3.07 \AA). However, a more recent study using scanning transmission electron microscopy (STEM) \cite{Wu2015_A} reported a value of 5.4 \AA\ for {\it b}/2, in excellent agreement with our result of 5.35(10) \AA\ for the sum $b_1+b_2+buckling$. This shows that despite the symmetry breaking in the direction [010] imposed by the surface, there are no large relaxations in perpendicular distances compared to the values usually obtained for the bulk. Because of the already weak interaction of van-der-Waals type, the surface presents basically the same distances between atoms from those found deeper in the material.

Nevertheless, we can analyze more carefully the contraction of the $b_1$ distance and the buckling. As already mentioned, within the phosphorene layer, each atom is covalently bonded to three neighbours, with two bonds connecting the nearest P atoms in the same plane (d$_1$), and the third bond conecting P atoms between the top and bottom of the phosphorene layer (d$_2$), see Fig. \ref{fig1}c. However, as reported by Morita \cite{Morita1986}, the in-plane distance is slightly smaller than the out-of-plane distance, d$_1$ = 2.222 \AA\ and d$_2$ = 2.277 \AA, respectively. Results reported by Brown {\it et al.} \cite{Brown1965} show a smaller difference between the distances: 2.224 \AA\ and 2.244 \AA, respectively. Our XPD results obtained were d$_1$ = 2.20(5) \AA\ and d$_2'$ = 2.18(8) \AA\ and d$_2''$ = 2.23(8) \AA. There are two different distances for $d_2$ due to the buckling. Thus, in the phosphorene surface layer, the average distance between neighboring atoms in different planes ($\sim$2.20 \AA) is very close to the distances between the neighboring atoms in the same plane (\mbox{2.20 \AA}), which indicates an uniformity in the size of the sp$^3$ bonds. Table 1 summarize the results presented.

\begin{table}[!h]
\centering
\caption{Lattice parameters, nearest neighbour distances and bond angles measured by {\it neutron powder diffraction} (NPD) \cite{Cartz1979}, {\it scanning tunneling microscopy} (STM) \cite{Zhang2009} and {\it photoelectron diffraction} (XPD) (this work). The parameter {\it b}/2 is defined as the sum $b_1+b_2+buckling$ for the XPD.}
\begin{tabular}{|c|c|c|c|} 
\hline
{\bf Parameter} & {\bf NPD\cite{Cartz1979}} & {\bf STM\cite{Zhang2009}} & {\bf XPD (this work)}\\ 
\hline
\hline
{\it a} (\AA) & 3.313 & 3.33 & 3.27\\
\hline
{\it b}/2 (\AA) & 5.2365 & - & 5.35\\
\hline
{\it c} (\AA) & 4.374 & 4.33 & 4.34\\
\hline
d$_1$ (\AA) & 2.222 & - & 2.20\\
\hline
d$_2$ (\AA) & 2.277 & - & 2.18(d$_2'$)/2.23(d$_2''$)\\
\hline
$\alpha_1$ (degree) & 96.5 & - & 96.1\\
\hline
$\alpha_2$ (degree) & 101.9 & - & 103.6($\alpha_2'$)/100.9($\alpha_2''$)\\
\hline
\end{tabular}\\
\end{table}

To summarize, the XPD results show that despite the perpendicular symmetry breaking imposed by the surface, the structure of the top phosphorene layer is very similar to that expected for the bulk, probably due to the already weak interaction between the bilayers. However, our results show that the top phosphorene layer in the black phosphorus is slightly displaced compared to its bulk structure \cite{Cartz1979} and presents a small contraction in the direction perpendicular to the surface. Furthermore, a small buckling among the surface atoms is observed, in agreement with results previously reported by STM \cite{Zhang2009}. The contraction of the surface phosphorene together with the buckling indicate an uniformity in the size of the sp$^3$ bonds between the P atoms on the surface.

\begin{acknowledgments}
This work received financial support from FAPESP, CNPq, and CAPES from Brazil. XPD measurements were done at LNLS under proposal PGM-19062. L.H.L. and L.B. thank CNPq for postdoctoral fellowship support. Authors thanks the LNLS staff, specially J. C. Cezar, for technical support during beamtime. 
\end{acknowledgments}

%\bibliography{bibliografia}

\begin{thebibliography}{40}%
\makeatletter
\providecommand \@ifxundefined [1]{%
 \@ifx{#1\undefined}
}%
\providecommand \@ifnum [1]{%
 \ifnum #1\expandafter \@firstoftwo
 \else \expandafter \@secondoftwo
 \fi
}%
\providecommand \@ifx [1]{%
 \ifx #1\expandafter \@firstoftwo
 \else \expandafter \@secondoftwo
 \fi
}%
\providecommand \natexlab [1]{#1}%
\providecommand \enquote  [1]{``#1''}%
\providecommand \bibnamefont  [1]{#1}%
\providecommand \bibfnamefont [1]{#1}%
\providecommand \citenamefont [1]{#1}%
\providecommand \href@noop [0]{\@secondoftwo}%
\providecommand \href [0]{\begingroup \@sanitize@url \@href}%
\providecommand \@href[1]{\@@startlink{#1}\@@href}%
\providecommand \@@href[1]{\endgroup#1\@@endlink}%
\providecommand \@sanitize@url [0]{\catcode `\\12\catcode `\$12\catcode
  `\&12\catcode `\#12\catcode `\^12\catcode `\_12\catcode `\%12\relax}%
\providecommand \@@startlink[1]{}%
\providecommand \@@endlink[0]{}%
\providecommand \url  [0]{\begingroup\@sanitize@url \@url }%
\providecommand \@url [1]{\endgroup\@href {#1}{\urlprefix }}%
\providecommand \urlprefix  [0]{URL }%
\providecommand \Eprint [0]{\href }%
\providecommand \doibase [0]{http://dx.doi.org/}%
\providecommand \selectlanguage [0]{\@gobble}%
\providecommand \bibinfo  [0]{\@secondoftwo}%
\providecommand \bibfield  [0]{\@secondoftwo}%
\providecommand \translation [1]{[#1]}%
\providecommand \BibitemOpen [0]{}%
\providecommand \bibitemStop [0]{}%
\providecommand \bibitemNoStop [0]{.\EOS\space}%
\providecommand \EOS [0]{\spacefactor3000\relax}%
\providecommand \BibitemShut  [1]{\csname bibitem#1\endcsname}%
\let\auto@bib@innerbib\@empty
%</preamble>
\bibitem [{\citenamefont {Novoselov}\ \emph {et~al.}(2004)\citenamefont
  {Novoselov}, \citenamefont {Geim}, \citenamefont {Morozov}, \citenamefont
  {Jiang}, \citenamefont {Zhang}, \citenamefont {Dubonos}, \citenamefont
  {Grigorieva},\ and\ \citenamefont {Firsov}}]{Novoselov2004}%
  \BibitemOpen
  \bibfield  {author} {\bibinfo {author} {\bibfnamefont {K.~S.}\ \bibnamefont
  {Novoselov}}, \bibinfo {author} {\bibfnamefont {A.~K.}\ \bibnamefont {Geim}},
  \bibinfo {author} {\bibfnamefont {S.~V.}\ \bibnamefont {Morozov}}, \bibinfo
  {author} {\bibfnamefont {D.}~\bibnamefont {Jiang}}, \bibinfo {author}
  {\bibfnamefont {Y.}~\bibnamefont {Zhang}}, \bibinfo {author} {\bibfnamefont
  {S.~V.}\ \bibnamefont {Dubonos}}, \bibinfo {author} {\bibfnamefont {I.~V.}\
  \bibnamefont {Grigorieva}}, \ and\ \bibinfo {author} {\bibfnamefont {A.~A.}\
  \bibnamefont {Firsov}},\ }\href@noop {} {\bibfield  {journal} {\bibinfo
  {journal} {Science}\ }\textbf {\bibinfo {volume} {306}},\ \bibinfo {pages}
  {666} (\bibinfo {year} {2004})}\BibitemShut {NoStop}%
\bibitem [{\citenamefont {Xu}\ \emph {et~al.}(2013)\citenamefont {Xu},
  \citenamefont {Liang}, \citenamefont {Shi},\ and\ \citenamefont
  {Chen}}]{Xu2013}%
  \BibitemOpen
  \bibfield  {author} {\bibinfo {author} {\bibfnamefont {M.}~\bibnamefont
  {Xu}}, \bibinfo {author} {\bibfnamefont {T.}~\bibnamefont {Liang}}, \bibinfo
  {author} {\bibfnamefont {M.}~\bibnamefont {Shi}}, \ and\ \bibinfo {author}
  {\bibfnamefont {H.}~\bibnamefont {Chen}},\ }\href@noop {} {\bibfield
  {journal} {\bibinfo  {journal} {Chem. Rev.}\ }\textbf {\bibinfo {volume}
  {113}},\ \bibinfo {pages} {3766} (\bibinfo {year} {2013})}\BibitemShut
  {NoStop}%
\bibitem [{\citenamefont {Geim}\ and\ \citenamefont
  {Grigorieva}(2013)}]{Geim2013}%
  \BibitemOpen
  \bibfield  {author} {\bibinfo {author} {\bibfnamefont {A.~K.}\ \bibnamefont
  {Geim}}\ and\ \bibinfo {author} {\bibfnamefont {I.~V.}\ \bibnamefont
  {Grigorieva}},\ }\href@noop {} {\bibfield  {journal} {\bibinfo  {journal}
  {Nature}\ }\textbf {\bibinfo {volume} {499}},\ \bibinfo {pages} {419}
  (\bibinfo {year} {2013})}\BibitemShut {NoStop}%
\bibitem [{\citenamefont {Ling}\ \emph {et~al.}(2015)\citenamefont {Ling},
  \citenamefont {Wang}, \citenamefont {Huang}, \citenamefont {Xia}, ,\ and\
  \citenamefont {Dresselhaus}}]{Ling2015}%
  \BibitemOpen
  \bibfield  {author} {\bibinfo {author} {\bibfnamefont {X.}~\bibnamefont
  {Ling}}, \bibinfo {author} {\bibfnamefont {H.}~\bibnamefont {Wang}}, \bibinfo
  {author} {\bibfnamefont {S.}~\bibnamefont {Huang}}, \bibinfo {author}
  {\bibfnamefont {F.}~\bibnamefont {Xia}}, , \ and\ \bibinfo {author}
  {\bibfnamefont {M.~S.}\ \bibnamefont {Dresselhaus}},\ }\href@noop {}
  {\bibfield  {journal} {\bibinfo  {journal} {Proc. of Nat. Acad. of Sci.}\
  }\textbf {\bibinfo {volume} {112}},\ \bibinfo {pages} {4523} (\bibinfo {year}
  {2015})}\BibitemShut {NoStop}%
\bibitem [{\citenamefont {Brown}\ and\ \citenamefont
  {Rundqvist}(1965)}]{Brown1965}%
  \BibitemOpen
  \bibfield  {author} {\bibinfo {author} {\bibfnamefont {A.}~\bibnamefont
  {Brown}}\ and\ \bibinfo {author} {\bibfnamefont {S.}~\bibnamefont
  {Rundqvist}},\ }\href@noop {} {\bibfield  {journal} {\bibinfo  {journal}
  {Acta Cryst.}\ }\textbf {\bibinfo {volume} {19}},\ \bibinfo {pages} {684}
  (\bibinfo {year} {1965})}\BibitemShut {NoStop}%
\bibitem [{\citenamefont {Morita}(1986)}]{Morita1986}%
  \BibitemOpen
  \bibfield  {author} {\bibinfo {author} {\bibfnamefont {A.}~\bibnamefont
  {Morita}},\ }\href@noop {} {\bibfield  {journal} {\bibinfo  {journal} {Appl.
  Phys. A}\ }\textbf {\bibinfo {volume} {39}},\ \bibinfo {pages} {227}
  (\bibinfo {year} {1986})}\BibitemShut {NoStop}%
\bibitem [{\citenamefont {Favron}\ \emph {et~al.}(2015)\citenamefont {Favron},
  \citenamefont {Gaufrès}, \citenamefont {Fossard}, \citenamefont
  {Phaneuf-L’Heureux}, \citenamefont {Tang}, \citenamefont {Lévesque},
  \citenamefont {Loiseau}, \citenamefont {Leonelli}, \citenamefont
  {Francoeur},\ and\ \citenamefont {Martel}}]{Favron2015}%
  \BibitemOpen
  \bibfield  {author} {\bibinfo {author} {\bibfnamefont {A.}~\bibnamefont
  {Favron}}, \bibinfo {author} {\bibfnamefont {E.}~\bibnamefont {Gaufrès}},
  \bibinfo {author} {\bibfnamefont {F.}~\bibnamefont {Fossard}}, \bibinfo
  {author} {\bibfnamefont {A.-L.}\ \bibnamefont {Phaneuf-L’Heureux}},
  \bibinfo {author} {\bibfnamefont {N.~Y.-W.}\ \bibnamefont {Tang}}, \bibinfo
  {author} {\bibfnamefont {P.~L.}\ \bibnamefont {Lévesque}}, \bibinfo {author}
  {\bibfnamefont {A.}~\bibnamefont {Loiseau}}, \bibinfo {author} {\bibfnamefont
  {R.}~\bibnamefont {Leonelli}}, \bibinfo {author} {\bibfnamefont
  {S.}~\bibnamefont {Francoeur}}, \ and\ \bibinfo {author} {\bibfnamefont
  {R.}~\bibnamefont {Martel}},\ }\href@noop {} {\bibfield  {journal} {\bibinfo
  {journal} {Nature Materials}\ }\textbf {\bibinfo {volume} {14}},\ \bibinfo
  {pages} {826} (\bibinfo {year} {2015})}\BibitemShut {NoStop}%
\bibitem [{\citenamefont {Tran}\ \emph {et~al.}(2014)\citenamefont {Tran},
  \citenamefont {Soklaski}, \citenamefont {Liang},\ and\ \citenamefont
  {Yang}}]{Tran2014}%
  \BibitemOpen
  \bibfield  {author} {\bibinfo {author} {\bibfnamefont {V.}~\bibnamefont
  {Tran}}, \bibinfo {author} {\bibfnamefont {R.}~\bibnamefont {Soklaski}},
  \bibinfo {author} {\bibfnamefont {Y.}~\bibnamefont {Liang}}, \ and\ \bibinfo
  {author} {\bibfnamefont {L.}~\bibnamefont {Yang}},\ }\href@noop {} {\bibfield
   {journal} {\bibinfo  {journal} {Phys. Rev. B}\ }\textbf {\bibinfo {volume}
  {89}},\ \bibinfo {pages} {235319} (\bibinfo {year} {2014})}\BibitemShut
  {NoStop}%
\bibitem [{\citenamefont {Asahina}\ \emph {et~al.}(1982)\citenamefont
  {Asahina}, \citenamefont {Shindo},\ and\ \citenamefont
  {Morita}}]{Asahina1982}%
  \BibitemOpen
  \bibfield  {author} {\bibinfo {author} {\bibfnamefont {H.}~\bibnamefont
  {Asahina}}, \bibinfo {author} {\bibfnamefont {K.}~\bibnamefont {Shindo}}, \
  and\ \bibinfo {author} {\bibfnamefont {A.}~\bibnamefont {Morita}},\
  }\href@noop {} {\bibfield  {journal} {\bibinfo  {journal} {J. Phys. Soc. J}\
  }\textbf {\bibinfo {volume} {51}},\ \bibinfo {pages} {1193} (\bibinfo {year}
  {1982})}\BibitemShut {NoStop}%
\bibitem [{\citenamefont {Goodman}\ \emph {et~al.}(1983)\citenamefont
  {Goodman}, \citenamefont {Ley},\ and\ \citenamefont {Bullett}}]{Goodman1983}%
  \BibitemOpen
  \bibfield  {author} {\bibinfo {author} {\bibfnamefont {N.~B.}\ \bibnamefont
  {Goodman}}, \bibinfo {author} {\bibfnamefont {L.}~\bibnamefont {Ley}}, \ and\
  \bibinfo {author} {\bibfnamefont {D.~W.}\ \bibnamefont {Bullett}},\
  }\href@noop {} {\bibfield  {journal} {\bibinfo  {journal} {Phys. Rev. B}\
  }\textbf {\bibinfo {volume} {27}},\ \bibinfo {pages} {7440} (\bibinfo {year}
  {1983})}\BibitemShut {NoStop}%
\bibitem [{\citenamefont {Li}\ \emph {et~al.}(2014{\natexlab{a}})\citenamefont
  {Li}, \citenamefont {Yu}, \citenamefont {Ye}, \citenamefont {Ge},
  \citenamefont {Ou}, \citenamefont {Wu}, \citenamefont {Feng}, \citenamefont
  {Chen},\ and\ \citenamefont {Zhang}}]{Li2014}%
  \BibitemOpen
  \bibfield  {author} {\bibinfo {author} {\bibfnamefont {L.}~\bibnamefont
  {Li}}, \bibinfo {author} {\bibfnamefont {Y.}~\bibnamefont {Yu}}, \bibinfo
  {author} {\bibfnamefont {G.~J.}\ \bibnamefont {Ye}}, \bibinfo {author}
  {\bibfnamefont {Q.}~\bibnamefont {Ge}}, \bibinfo {author} {\bibfnamefont
  {X.}~\bibnamefont {Ou}}, \bibinfo {author} {\bibfnamefont {H.}~\bibnamefont
  {Wu}}, \bibinfo {author} {\bibfnamefont {D.}~\bibnamefont {Feng}}, \bibinfo
  {author} {\bibfnamefont {X.~H.}\ \bibnamefont {Chen}}, \ and\ \bibinfo
  {author} {\bibfnamefont {Y.}~\bibnamefont {Zhang}},\ }\href@noop {}
  {\bibfield  {journal} {\bibinfo  {journal} {Nat. Nano.}\ }\textbf {\bibinfo
  {volume} {9}},\ \bibinfo {pages} {372} (\bibinfo {year}
  {2014}{\natexlab{a}})}\BibitemShut {NoStop}%
\bibitem [{\citenamefont {Du}\ \emph {et~al.}(2010)\citenamefont {Du},
  \citenamefont {Ouyang}, \citenamefont {Shi},\ and\ \citenamefont
  {Lei}}]{Du2010}%
  \BibitemOpen
  \bibfield  {author} {\bibinfo {author} {\bibfnamefont {Y.}~\bibnamefont
  {Du}}, \bibinfo {author} {\bibfnamefont {C.}~\bibnamefont {Ouyang}}, \bibinfo
  {author} {\bibfnamefont {S.}~\bibnamefont {Shi}}, \ and\ \bibinfo {author}
  {\bibfnamefont {M.}~\bibnamefont {Lei}},\ }\href@noop {} {\bibfield
  {journal} {\bibinfo  {journal} {J. Appl. Phys.}\ }\textbf {\bibinfo {volume}
  {107}},\ \bibinfo {pages} {093718} (\bibinfo {year} {2010})}\BibitemShut
  {NoStop}%
\bibitem [{\citenamefont {Takahashi}\ \emph {et~al.}(1984)\citenamefont
  {Takahashi}, \citenamefont {Tokailin}, \citenamefont {Suzuki}, \citenamefont
  {Sagawa},\ and\ \citenamefont {Shirotani}}]{Takahashi1984}%
  \BibitemOpen
  \bibfield  {author} {\bibinfo {author} {\bibfnamefont {T.}~\bibnamefont
  {Takahashi}}, \bibinfo {author} {\bibfnamefont {H.}~\bibnamefont {Tokailin}},
  \bibinfo {author} {\bibfnamefont {S.}~\bibnamefont {Suzuki}}, \bibinfo
  {author} {\bibfnamefont {T.}~\bibnamefont {Sagawa}}, \ and\ \bibinfo {author}
  {\bibfnamefont {I.}~\bibnamefont {Shirotani}},\ }\href@noop {} {\bibfield
  {journal} {\bibinfo  {journal} {Phys. Rev. B}\ }\textbf {\bibinfo {volume}
  {29}},\ \bibinfo {pages} {1105} (\bibinfo {year} {1984})}\BibitemShut
  {NoStop}%
\bibitem [{\citenamefont {Takahashi}\ \emph {et~al.}(1986)\citenamefont
  {Takahashi}, \citenamefont {Gunasekara}, \citenamefont {Ohsawa},
  \citenamefont {Ishii}, \citenamefont {Kinoshita}, \citenamefont {Suzuki},
  \citenamefont {Sagawa}, \citenamefont {Kato}, \citenamefont {Miyahara},\ and\
  \citenamefont {Shirotani}}]{Takahashi1986}%
  \BibitemOpen
  \bibfield  {author} {\bibinfo {author} {\bibfnamefont {T.}~\bibnamefont
  {Takahashi}}, \bibinfo {author} {\bibfnamefont {N.}~\bibnamefont
  {Gunasekara}}, \bibinfo {author} {\bibfnamefont {H.}~\bibnamefont {Ohsawa}},
  \bibinfo {author} {\bibfnamefont {H.}~\bibnamefont {Ishii}}, \bibinfo
  {author} {\bibfnamefont {T.}~\bibnamefont {Kinoshita}}, \bibinfo {author}
  {\bibfnamefont {S.}~\bibnamefont {Suzuki}}, \bibinfo {author} {\bibfnamefont
  {T.}~\bibnamefont {Sagawa}}, \bibinfo {author} {\bibfnamefont
  {H.}~\bibnamefont {Kato}}, \bibinfo {author} {\bibfnamefont {T.}~\bibnamefont
  {Miyahara}}, \ and\ \bibinfo {author} {\bibfnamefont {I.}~\bibnamefont
  {Shirotani}},\ }\href@noop {} {\bibfield  {journal} {\bibinfo  {journal}
  {Phys. Rev. B}\ }\textbf {\bibinfo {volume} {33}},\ \bibinfo {pages} {4324}
  (\bibinfo {year} {1986})}\BibitemShut {NoStop}%
\bibitem [{\citenamefont {Han}\ \emph {et~al.}(2014)\citenamefont {Han},
  \citenamefont {Yao}, \citenamefont {Bai}, \citenamefont {Miao}, \citenamefont
  {Zhu}, \citenamefont {Guan}, \citenamefont {Wang}, \citenamefont {Gao},
  \citenamefont {Liu}, \citenamefont {Qian}, \citenamefont {Liu},\ and\
  \citenamefont {feng Jia}}]{Han2014}%
  \BibitemOpen
  \bibfield  {author} {\bibinfo {author} {\bibfnamefont {C.~Q.}\ \bibnamefont
  {Han}}, \bibinfo {author} {\bibfnamefont {M.~Y.}\ \bibnamefont {Yao}},
  \bibinfo {author} {\bibfnamefont {X.~X.}\ \bibnamefont {Bai}}, \bibinfo
  {author} {\bibfnamefont {L.}~\bibnamefont {Miao}}, \bibinfo {author}
  {\bibfnamefont {F.}~\bibnamefont {Zhu}}, \bibinfo {author} {\bibfnamefont
  {D.~D.}\ \bibnamefont {Guan}}, \bibinfo {author} {\bibfnamefont
  {S.}~\bibnamefont {Wang}}, \bibinfo {author} {\bibfnamefont {C.~L.}\
  \bibnamefont {Gao}}, \bibinfo {author} {\bibfnamefont {C.}~\bibnamefont
  {Liu}}, \bibinfo {author} {\bibfnamefont {D.}~\bibnamefont {Qian}}, \bibinfo
  {author} {\bibfnamefont {Y.}~\bibnamefont {Liu}}, \ and\ \bibinfo {author}
  {\bibfnamefont {J.}~\bibnamefont {feng Jia}},\ }\href@noop {} {\bibfield
  {journal} {\bibinfo  {journal} {Phys. Rev. B}\ }\textbf {\bibinfo {volume}
  {90}},\ \bibinfo {pages} {085101} (\bibinfo {year} {2014})}\BibitemShut
  {NoStop}%
\bibitem [{\citenamefont {Rodin}\ \emph {et~al.}(2014)\citenamefont {Rodin},
  \citenamefont {Carvalho},\ and\ \citenamefont {Neto}}]{Rodin2014}%
  \BibitemOpen
  \bibfield  {author} {\bibinfo {author} {\bibfnamefont {A.~S.}\ \bibnamefont
  {Rodin}}, \bibinfo {author} {\bibfnamefont {A.}~\bibnamefont {Carvalho}}, \
  and\ \bibinfo {author} {\bibfnamefont {A.~H.~C.}\ \bibnamefont {Neto}},\
  }\href@noop {} {\bibfield  {journal} {\bibinfo  {journal} {Phys. Rev. Lett.}\
  }\textbf {\bibinfo {volume} {112}},\ \bibinfo {pages} {176801} (\bibinfo
  {year} {2014})}\BibitemShut {NoStop}%
\bibitem [{\citenamefont {Li}\ \emph {et~al.}(2014{\natexlab{b}})\citenamefont
  {Li}, \citenamefont {Yang},\ and\ \citenamefont {Li}}]{Li2014_A}%
  \BibitemOpen
  \bibfield  {author} {\bibinfo {author} {\bibfnamefont {Y.}~\bibnamefont
  {Li}}, \bibinfo {author} {\bibfnamefont {S.}~\bibnamefont {Yang}}, \ and\
  \bibinfo {author} {\bibfnamefont {J.}~\bibnamefont {Li}},\ }\href@noop {}
  {\bibfield  {journal} {\bibinfo  {journal} {J. Phys. Chem. C}\ }\textbf
  {\bibinfo {volume} {118}},\ \bibinfo {pages} {23970} (\bibinfo {year}
  {2014}{\natexlab{b}})}\BibitemShut {NoStop}%
\bibitem [{\citenamefont {Peng}\ \emph {et~al.}(2014)\citenamefont {Peng},
  \citenamefont {Wei},\ and\ \citenamefont {Copple}}]{Peng2014}%
  \BibitemOpen
  \bibfield  {author} {\bibinfo {author} {\bibfnamefont {X.}~\bibnamefont
  {Peng}}, \bibinfo {author} {\bibfnamefont {Q.}~\bibnamefont {Wei}}, \ and\
  \bibinfo {author} {\bibfnamefont {A.}~\bibnamefont {Copple}},\ }\href@noop {}
  {\bibfield  {journal} {\bibinfo  {journal} {Phys. Rev. B}\ }\textbf {\bibinfo
  {volume} {90}},\ \bibinfo {pages} {085402} (\bibinfo {year}
  {2014})}\BibitemShut {NoStop}%
\bibitem [{\citenamefont {Dai}\ and\ \citenamefont {Zeng}(2014)}]{Dai2014}%
  \BibitemOpen
  \bibfield  {author} {\bibinfo {author} {\bibfnamefont {J.}~\bibnamefont
  {Dai}}\ and\ \bibinfo {author} {\bibfnamefont {X.~C.}\ \bibnamefont {Zeng}},\
  }\href@noop {} {\bibfield  {journal} {\bibinfo  {journal} {J. Phys. Chem.
  Lett.}\ }\textbf {\bibinfo {volume} {5}},\ \bibinfo {pages} {1289} (\bibinfo
  {year} {2014})}\BibitemShut {NoStop}%
\bibitem [{\citenamefont {Guo}\ \emph {et~al.}(2014)\citenamefont {Guo},
  \citenamefont {Lu}, \citenamefont {Dai}, \citenamefont {Wu},\ and\
  \citenamefont {Zeng}}]{Guo2014}%
  \BibitemOpen
  \bibfield  {author} {\bibinfo {author} {\bibfnamefont {H.}~\bibnamefont
  {Guo}}, \bibinfo {author} {\bibfnamefont {N.}~\bibnamefont {Lu}}, \bibinfo
  {author} {\bibfnamefont {J.}~\bibnamefont {Dai}}, \bibinfo {author}
  {\bibfnamefont {X.}~\bibnamefont {Wu}}, \ and\ \bibinfo {author}
  {\bibfnamefont {X.~C.}\ \bibnamefont {Zeng}},\ }\href@noop {} {\bibfield
  {journal} {\bibinfo  {journal} {J. Phys. Chem. C}\ }\textbf {\bibinfo
  {volume} {118}},\ \bibinfo {pages} {14051} (\bibinfo {year}
  {2014})}\BibitemShut {NoStop}%
\bibitem [{\citenamefont {Liu}\ \emph {et~al.}(2015)\citenamefont {Liu},
  \citenamefont {Zhang}, \citenamefont {Abdalla}, \citenamefont {Fazzio},\ and\
  \citenamefont {Zunger}}]{Liu2015}%
  \BibitemOpen
  \bibfield  {author} {\bibinfo {author} {\bibfnamefont {Q.}~\bibnamefont
  {Liu}}, \bibinfo {author} {\bibfnamefont {X.}~\bibnamefont {Zhang}}, \bibinfo
  {author} {\bibfnamefont {L.~B.}\ \bibnamefont {Abdalla}}, \bibinfo {author}
  {\bibfnamefont {A.}~\bibnamefont {Fazzio}}, \ and\ \bibinfo {author}
  {\bibfnamefont {A.}~\bibnamefont {Zunger}},\ }\href@noop {} {\bibfield
  {journal} {\bibinfo  {journal} {Nano Lett.}\ }\textbf {\bibinfo {volume}
  {15}},\ \bibinfo {pages} {1222} (\bibinfo {year} {2015})}\BibitemShut
  {NoStop}%
\bibitem [{\citenamefont {Wu}\ \emph {et~al.}(2015{\natexlab{a}})\citenamefont
  {Wu}, \citenamefont {Shen}, \citenamefont {Yang}, \citenamefont {Cai},
  \citenamefont {Huang},\ and\ \citenamefont {Feng}}]{Wu2015}%
  \BibitemOpen
  \bibfield  {author} {\bibinfo {author} {\bibfnamefont {Q.}~\bibnamefont
  {Wu}}, \bibinfo {author} {\bibfnamefont {L.}~\bibnamefont {Shen}}, \bibinfo
  {author} {\bibfnamefont {M.}~\bibnamefont {Yang}}, \bibinfo {author}
  {\bibfnamefont {Y.}~\bibnamefont {Cai}}, \bibinfo {author} {\bibfnamefont
  {Z.}~\bibnamefont {Huang}}, \ and\ \bibinfo {author} {\bibfnamefont {Y.~P.}\
  \bibnamefont {Feng}},\ }\href@noop {} {\bibfield  {journal} {\bibinfo
  {journal} {Phys. Rev. B}\ }\textbf {\bibinfo {volume} {92}},\ \bibinfo
  {pages} {035436} (\bibinfo {year} {2015}{\natexlab{a}})}\BibitemShut
  {NoStop}%
\bibitem [{\citenamefont {Fei}\ \emph {et~al.}(2015)\citenamefont {Fei},
  \citenamefont {Tran},\ and\ \citenamefont {Yang}}]{Fei2015}%
  \BibitemOpen
  \bibfield  {author} {\bibinfo {author} {\bibfnamefont {R.}~\bibnamefont
  {Fei}}, \bibinfo {author} {\bibfnamefont {V.}~\bibnamefont {Tran}}, \ and\
  \bibinfo {author} {\bibfnamefont {L.}~\bibnamefont {Yang}},\ }\href@noop {}
  {\bibfield  {journal} {\bibinfo  {journal} {Phys. Rev. B}\ }\textbf {\bibinfo
  {volume} {91}},\ \bibinfo {pages} {195319} (\bibinfo {year}
  {2015})}\BibitemShut {NoStop}%
\bibitem [{\citenamefont {Kim}\ \emph {et~al.}(2015)\citenamefont {Kim},
  \citenamefont {Baik}, \citenamefont {Ryu}, \citenamefont {Sohn},
  \citenamefont {Park}, \citenamefont {Park}, \citenamefont {Denlinger},
  \citenamefont {Yi}, \citenamefont {Choi},\ and\ \citenamefont
  {Kim}}]{Kim2015}%
  \BibitemOpen
  \bibfield  {author} {\bibinfo {author} {\bibfnamefont {J.}~\bibnamefont
  {Kim}}, \bibinfo {author} {\bibfnamefont {S.~S.}\ \bibnamefont {Baik}},
  \bibinfo {author} {\bibfnamefont {S.~H.}\ \bibnamefont {Ryu}}, \bibinfo
  {author} {\bibfnamefont {Y.}~\bibnamefont {Sohn}}, \bibinfo {author}
  {\bibfnamefont {S.}~\bibnamefont {Park}}, \bibinfo {author} {\bibfnamefont
  {B.-G.}\ \bibnamefont {Park}}, \bibinfo {author} {\bibfnamefont
  {J.}~\bibnamefont {Denlinger}}, \bibinfo {author} {\bibfnamefont
  {Y.}~\bibnamefont {Yi}}, \bibinfo {author} {\bibfnamefont {H.~J.}\
  \bibnamefont {Choi}}, \ and\ \bibinfo {author} {\bibfnamefont {K.~S.}\
  \bibnamefont {Kim}},\ }\href@noop {} {\bibfield  {journal} {\bibinfo
  {journal} {Science}\ }\textbf {\bibinfo {volume} {349}},\ \bibinfo {pages}
  {723} (\bibinfo {year} {2015})}\BibitemShut {NoStop}%
\bibitem [{\citenamefont {Ohta}\ \emph {et~al.}(2006)\citenamefont {Ohta},
  \citenamefont {Bostwick}, \citenamefont {Seyller}, \citenamefont {Horn},\
  and\ \citenamefont {Rotenberg}}]{Ohta2006}%
  \BibitemOpen
  \bibfield  {author} {\bibinfo {author} {\bibfnamefont {T.}~\bibnamefont
  {Ohta}}, \bibinfo {author} {\bibfnamefont {A.}~\bibnamefont {Bostwick}},
  \bibinfo {author} {\bibfnamefont {T.}~\bibnamefont {Seyller}}, \bibinfo
  {author} {\bibfnamefont {K.}~\bibnamefont {Horn}}, \ and\ \bibinfo {author}
  {\bibfnamefont {E.}~\bibnamefont {Rotenberg}},\ }\href@noop {} {\bibfield
  {journal} {\bibinfo  {journal} {Science}\ }\textbf {\bibinfo {volume}
  {313}},\ \bibinfo {pages} {951} (\bibinfo {year} {2006})}\BibitemShut
  {NoStop}%
\bibitem [{\citenamefont {Cartz}\ \emph {et~al.}(1979)\citenamefont {Cartz},
  \citenamefont {Srinivasa}, \citenamefont {Riedner}, \citenamefont
  {Jorgensen},\ and\ \citenamefont {Worlton}}]{Cartz1979}%
  \BibitemOpen
  \bibfield  {author} {\bibinfo {author} {\bibfnamefont {L.}~\bibnamefont
  {Cartz}}, \bibinfo {author} {\bibfnamefont {S.~R.}\ \bibnamefont
  {Srinivasa}}, \bibinfo {author} {\bibfnamefont {R.~J.}\ \bibnamefont
  {Riedner}}, \bibinfo {author} {\bibfnamefont {J.~D.}\ \bibnamefont
  {Jorgensen}}, \ and\ \bibinfo {author} {\bibfnamefont {T.~G.}\ \bibnamefont
  {Worlton}},\ }\href@noop {} {\bibfield  {journal} {\bibinfo  {journal} {J.
  Chem. Phys.}\ }\textbf {\bibinfo {volume} {71}},\ \bibinfo {pages} {1718}
  (\bibinfo {year} {1979})}\BibitemShut {NoStop}%
\bibitem [{\citenamefont {Wu}\ \emph {et~al.}(2015{\natexlab{b}})\citenamefont
  {Wu}, \citenamefont {Topsakal}, \citenamefont {Low}, \citenamefont {Robbins},
  \citenamefont {Haratipour}, \citenamefont {Jeong}, \citenamefont
  {Wentzcovitch}, \citenamefont {Koester},\ and\ \citenamefont
  {Mkhoyan}}]{Wu2015_A}%
  \BibitemOpen
  \bibfield  {author} {\bibinfo {author} {\bibfnamefont {R.~J.}\ \bibnamefont
  {Wu}}, \bibinfo {author} {\bibfnamefont {M.}~\bibnamefont {Topsakal}},
  \bibinfo {author} {\bibfnamefont {T.}~\bibnamefont {Low}}, \bibinfo {author}
  {\bibfnamefont {M.~C.}\ \bibnamefont {Robbins}}, \bibinfo {author}
  {\bibfnamefont {N.}~\bibnamefont {Haratipour}}, \bibinfo {author}
  {\bibfnamefont {J.~S.}\ \bibnamefont {Jeong}}, \bibinfo {author}
  {\bibfnamefont {R.~M.}\ \bibnamefont {Wentzcovitch}}, \bibinfo {author}
  {\bibfnamefont {S.~J.}\ \bibnamefont {Koester}}, \ and\ \bibinfo {author}
  {\bibfnamefont {K.~A.}\ \bibnamefont {Mkhoyan}},\ }\href@noop {} {\bibfield
  {journal} {\bibinfo  {journal} {J. Vac. Sci. Technol. A}\ }\textbf {\bibinfo
  {volume} {33}},\ \bibinfo {pages} {060604} (\bibinfo {year}
  {2015}{\natexlab{b}})}\BibitemShut {NoStop}%
\bibitem [{\citenamefont {Zhang}\ \emph {et~al.}(2009)\citenamefont {Zhang},
  \citenamefont {J.~C.~Lian}, \citenamefont {Jiang}, \citenamefont {Liu},
  \citenamefont {Hu}, \citenamefont {Xiao}, \citenamefont {Du}, \citenamefont
  {Sun},\ and\ \citenamefont {Gao}}]{Zhang2009}%
  \BibitemOpen
  \bibfield  {author} {\bibinfo {author} {\bibfnamefont {C.~D.}\ \bibnamefont
  {Zhang}}, \bibinfo {author} {\bibfnamefont {W.~Y.}\ \bibnamefont
  {J.~C.~Lian}}, \bibinfo {author} {\bibfnamefont {Y.~H.}\ \bibnamefont
  {Jiang}}, \bibinfo {author} {\bibfnamefont {L.~W.}\ \bibnamefont {Liu}},
  \bibinfo {author} {\bibfnamefont {H.}~\bibnamefont {Hu}}, \bibinfo {author}
  {\bibfnamefont {W.~D.}\ \bibnamefont {Xiao}}, \bibinfo {author}
  {\bibfnamefont {S.~X.}\ \bibnamefont {Du}}, \bibinfo {author} {\bibfnamefont
  {L.~L.}\ \bibnamefont {Sun}}, \ and\ \bibinfo {author} {\bibfnamefont
  {H.~J.}\ \bibnamefont {Gao}},\ }\href@noop {} {\bibfield  {journal} {\bibinfo
   {journal} {J. Phys. Chem. C}\ }\textbf {\bibinfo {volume} {113}},\ \bibinfo
  {pages} {18823} (\bibinfo {year} {2009})}\BibitemShut {NoStop}%
\bibitem [{\citenamefont {Aierken}\ \emph {et~al.}(2015)\citenamefont
  {Aierken}, \citenamefont {Çakir}, \citenamefont {Sevik},\ and\ \citenamefont
  {Peeters}}]{Aierken2015}%
  \BibitemOpen
  \bibfield  {author} {\bibinfo {author} {\bibfnamefont {Y.}~\bibnamefont
  {Aierken}}, \bibinfo {author} {\bibfnamefont {D.}~\bibnamefont {Çakir}},
  \bibinfo {author} {\bibfnamefont {C.}~\bibnamefont {Sevik}}, \ and\ \bibinfo
  {author} {\bibfnamefont {F.~M.}\ \bibnamefont {Peeters}},\ }\href@noop {}
  {\bibfield  {journal} {\bibinfo  {journal} {Phys. Rev. B}\ }\textbf {\bibinfo
  {volume} {92}},\ \bibinfo {pages} {081408} (\bibinfo {year}
  {2015})}\BibitemShut {NoStop}%
\bibitem [{\citenamefont {Edmonds}\ \emph {et~al.}(2015)\citenamefont
  {Edmonds}, \citenamefont {Tadich}, \citenamefont {Carvalho}, \citenamefont
  {Ziletti}, \citenamefont {O’Donnell}, \citenamefont {Koenig}, \citenamefont
  {Coker}, \citenamefont {Ozyilmaz}, \citenamefont {Neto},\ and\ \citenamefont
  {Fuhrer}}]{Edmonds2015}%
  \BibitemOpen
  \bibfield  {author} {\bibinfo {author} {\bibfnamefont {M.~T.}\ \bibnamefont
  {Edmonds}}, \bibinfo {author} {\bibfnamefont {A.}~\bibnamefont {Tadich}},
  \bibinfo {author} {\bibfnamefont {A.}~\bibnamefont {Carvalho}}, \bibinfo
  {author} {\bibfnamefont {A.}~\bibnamefont {Ziletti}}, \bibinfo {author}
  {\bibfnamefont {K.~M.}\ \bibnamefont {O’Donnell}}, \bibinfo {author}
  {\bibfnamefont {S.~P.}\ \bibnamefont {Koenig}}, \bibinfo {author}
  {\bibfnamefont {D.~F.}\ \bibnamefont {Coker}}, \bibinfo {author}
  {\bibfnamefont {B.}~\bibnamefont {Ozyilmaz}}, \bibinfo {author}
  {\bibfnamefont {A.~H.~C.}\ \bibnamefont {Neto}}, \ and\ \bibinfo {author}
  {\bibfnamefont {M.~S.}\ \bibnamefont {Fuhrer}},\ }\href@noop {} {\bibfield
  {journal} {\bibinfo  {journal} {ACS Appl. Mater. Interf.}\ }\textbf {\bibinfo
  {volume} {7}},\ \bibinfo {pages} {14557} (\bibinfo {year}
  {2015})}\BibitemShut {NoStop}%
\bibitem [{\citenamefont {Xia}\ \emph {et~al.}(2014)\citenamefont {Xia},
  \citenamefont {Wang},\ and\ \citenamefont {Jia}}]{Xia2014}%
  \BibitemOpen
  \bibfield  {author} {\bibinfo {author} {\bibfnamefont {F.}~\bibnamefont
  {Xia}}, \bibinfo {author} {\bibfnamefont {H.}~\bibnamefont {Wang}}, \ and\
  \bibinfo {author} {\bibfnamefont {Y.}~\bibnamefont {Jia}},\ }\href@noop {}
  {\bibfield  {journal} {\bibinfo  {journal} {Nat. Commun.}\ }\textbf {\bibinfo
  {volume} {5}},\ \bibinfo {pages} {4458} (\bibinfo {year} {2014})}\BibitemShut
  {NoStop}%
\bibitem [{\citenamefont {Cezar}\ \emph {et~al.}(2013)\citenamefont {Cezar},
  \citenamefont {Fonseca}, \citenamefont {Rodrigues}, \citenamefont
  {de~Castro}, \citenamefont {Neuenschwander}, \citenamefont {Rodrigues},
  \citenamefont {Meyer}, \citenamefont {Ribeiro}, \citenamefont {Moreira},
  \citenamefont {Piton}, \citenamefont {Raulik}, \citenamefont {Donadio},
  \citenamefont {Seraphim}, \citenamefont {Barbosa}, \citenamefont {de~Siervo},
  \citenamefont {Landers},\ and\ \citenamefont {de~Brito}}]{Cezar2013}%
  \BibitemOpen
  \bibfield  {author} {\bibinfo {author} {\bibfnamefont {J.~C.}\ \bibnamefont
  {Cezar}}, \bibinfo {author} {\bibfnamefont {P.~T.}\ \bibnamefont {Fonseca}},
  \bibinfo {author} {\bibfnamefont {G.~L. M.~P.}\ \bibnamefont {Rodrigues}},
  \bibinfo {author} {\bibfnamefont {A.~R.~B.}\ \bibnamefont {de~Castro}},
  \bibinfo {author} {\bibfnamefont {R.~T.}\ \bibnamefont {Neuenschwander}},
  \bibinfo {author} {\bibfnamefont {F.}~\bibnamefont {Rodrigues}}, \bibinfo
  {author} {\bibfnamefont {B.~C.}\ \bibnamefont {Meyer}}, \bibinfo {author}
  {\bibfnamefont {L.~F.~S.}\ \bibnamefont {Ribeiro}}, \bibinfo {author}
  {\bibfnamefont {A.~F. A.~G.}\ \bibnamefont {Moreira}}, \bibinfo {author}
  {\bibfnamefont {J.~R.}\ \bibnamefont {Piton}}, \bibinfo {author}
  {\bibfnamefont {M.~A.}\ \bibnamefont {Raulik}}, \bibinfo {author}
  {\bibfnamefont {M.~P.}\ \bibnamefont {Donadio}}, \bibinfo {author}
  {\bibfnamefont {R.~M.}\ \bibnamefont {Seraphim}}, \bibinfo {author}
  {\bibfnamefont {M.~A.}\ \bibnamefont {Barbosa}}, \bibinfo {author}
  {\bibfnamefont {A.}~\bibnamefont {de~Siervo}}, \bibinfo {author}
  {\bibfnamefont {R.}~\bibnamefont {Landers}}, \ and\ \bibinfo {author}
  {\bibfnamefont {A.~N.}\ \bibnamefont {de~Brito}},\ }\href@noop {} {\bibfield
  {journal} {\bibinfo  {journal} {J. of Phys.: Confer. Series}\ }\textbf
  {\bibinfo {volume} {425}},\ \bibinfo {pages} {072015} (\bibinfo {year}
  {2013})}\BibitemShut {NoStop}%
\bibitem [{\citenamefont {Chen}\ \emph {et~al.}(1998)\citenamefont {Chen},
  \citenamefont {de~Abajo}, \citenamefont {Chasse}, \citenamefont {Ynzunza},
  \citenamefont {Kaduwela}, \citenamefont {Hove},\ and\ \citenamefont
  {Fadley}}]{Chen1998}%
  \BibitemOpen
  \bibfield  {author} {\bibinfo {author} {\bibfnamefont {Y.}~\bibnamefont
  {Chen}}, \bibinfo {author} {\bibfnamefont {F.~J.~G.}\ \bibnamefont
  {de~Abajo}}, \bibinfo {author} {\bibfnamefont {A.}~\bibnamefont {Chasse}},
  \bibinfo {author} {\bibfnamefont {R.~X.}\ \bibnamefont {Ynzunza}}, \bibinfo
  {author} {\bibfnamefont {A.~P.}\ \bibnamefont {Kaduwela}}, \bibinfo {author}
  {\bibfnamefont {M.~A.~V.}\ \bibnamefont {Hove}}, \ and\ \bibinfo {author}
  {\bibfnamefont {C.~S.}\ \bibnamefont {Fadley}},\ }\href@noop {} {\bibfield
  {journal} {\bibinfo  {journal} {Phys. Rev. B}\ }\textbf {\bibinfo {volume}
  {58}},\ \bibinfo {pages} {13121} (\bibinfo {year} {1998})}\BibitemShut
  {NoStop}%
\bibitem [{\citenamefont {Viana}\ \emph {et~al.}(2007)\citenamefont {Viana},
  \citenamefont {Muino}, \citenamefont {Soares}, \citenamefont {Hove},\ and\
  \citenamefont {de~Carvalho}}]{Viana2007}%
  \BibitemOpen
  \bibfield  {author} {\bibinfo {author} {\bibfnamefont {M.~L.}\ \bibnamefont
  {Viana}}, \bibinfo {author} {\bibfnamefont {R.~D.}\ \bibnamefont {Muino}},
  \bibinfo {author} {\bibfnamefont {E.~A.}\ \bibnamefont {Soares}}, \bibinfo
  {author} {\bibfnamefont {M.~A.~V.}\ \bibnamefont {Hove}}, \ and\ \bibinfo
  {author} {\bibfnamefont {V.~E.}\ \bibnamefont {de~Carvalho}},\ }\href@noop {}
  {\bibfield  {journal} {\bibinfo  {journal} {J. Phys. Condens. Matter}\
  }\textbf {\bibinfo {volume} {19}},\ \bibinfo {pages} {446002} (\bibinfo
  {year} {2007})}\BibitemShut {NoStop}%
\bibitem [{\citenamefont {de~Siervo}\ \emph {et~al.}(2002)\citenamefont
  {de~Siervo}, \citenamefont {Soares}, \citenamefont {Landers}, \citenamefont
  {Fazan}, \citenamefont {Morais},\ and\ \citenamefont
  {Kleiman}}]{deSiervo2002}%
  \BibitemOpen
  \bibfield  {author} {\bibinfo {author} {\bibfnamefont {A.}~\bibnamefont
  {de~Siervo}}, \bibinfo {author} {\bibfnamefont {E.~A.}\ \bibnamefont
  {Soares}}, \bibinfo {author} {\bibfnamefont {R.}~\bibnamefont {Landers}},
  \bibinfo {author} {\bibfnamefont {T.~A.}\ \bibnamefont {Fazan}}, \bibinfo
  {author} {\bibfnamefont {J.}~\bibnamefont {Morais}}, \ and\ \bibinfo {author}
  {\bibfnamefont {G.~G.}\ \bibnamefont {Kleiman}},\ }\href@noop {} {\bibfield
  {journal} {\bibinfo  {journal} {Surface Science}\ }\textbf {\bibinfo {volume}
  {504}},\ \bibinfo {pages} {215} (\bibinfo {year} {2002})}\BibitemShut
  {NoStop}%
\bibitem [{\citenamefont {Booth}\ \emph {et~al.}(1997)\citenamefont {Booth},
  \citenamefont {Davis}, \citenamefont {Toomes}, \citenamefont {Woodruff},
  \citenamefont {Hirschmugl}, \citenamefont {Schindler}, \citenamefont
  {Schaff}, \citenamefont {Fernandez}, \citenamefont {Theobald}, \citenamefont
  {Hofmann}, \citenamefont {Lindsay}, \citenamefont {Giessel}, \citenamefont
  {Baumgaertel},\ and\ \citenamefont {Bradshaw}}]{Booth1997}%
  \BibitemOpen
  \bibfield  {author} {\bibinfo {author} {\bibfnamefont {N.~A.}\ \bibnamefont
  {Booth}}, \bibinfo {author} {\bibfnamefont {R.}~\bibnamefont {Davis}},
  \bibinfo {author} {\bibfnamefont {R.}~\bibnamefont {Toomes}}, \bibinfo
  {author} {\bibfnamefont {D.~P.}\ \bibnamefont {Woodruff}}, \bibinfo {author}
  {\bibfnamefont {C.}~\bibnamefont {Hirschmugl}}, \bibinfo {author}
  {\bibfnamefont {K.~M.}\ \bibnamefont {Schindler}}, \bibinfo {author}
  {\bibfnamefont {O.}~\bibnamefont {Schaff}}, \bibinfo {author} {\bibfnamefont
  {V.}~\bibnamefont {Fernandez}}, \bibinfo {author} {\bibfnamefont
  {A.}~\bibnamefont {Theobald}}, \bibinfo {author} {\bibfnamefont
  {P.}~\bibnamefont {Hofmann}}, \bibinfo {author} {\bibfnamefont
  {R.}~\bibnamefont {Lindsay}}, \bibinfo {author} {\bibfnamefont
  {T.}~\bibnamefont {Giessel}}, \bibinfo {author} {\bibfnamefont
  {P.}~\bibnamefont {Baumgaertel}}, \ and\ \bibinfo {author} {\bibfnamefont
  {A.~M.}\ \bibnamefont {Bradshaw}},\ }\href@noop {} {\bibfield  {journal}
  {\bibinfo  {journal} {Surf. Sci.}\ }\textbf {\bibinfo {volume} {387}},\
  \bibinfo {pages} {152} (\bibinfo {year} {1997})}\BibitemShut {NoStop}%
\bibitem [{\citenamefont {Bondino}\ \emph {et~al.}(2002)\citenamefont
  {Bondino}, \citenamefont {Comelli}, \citenamefont {A.Baraldi}, \citenamefont
  {Rosei}, \citenamefont {Lizzit}, \citenamefont {Goldoni}, \citenamefont
  {Larciprete},\ and\ \citenamefont {Paolucci}}]{Bondino2002}%
  \BibitemOpen
  \bibfield  {author} {\bibinfo {author} {\bibfnamefont {F.}~\bibnamefont
  {Bondino}}, \bibinfo {author} {\bibfnamefont {G.}~\bibnamefont {Comelli}},
  \bibinfo {author} {\bibnamefont {A.Baraldi}}, \bibinfo {author}
  {\bibfnamefont {R.}~\bibnamefont {Rosei}}, \bibinfo {author} {\bibfnamefont
  {S.}~\bibnamefont {Lizzit}}, \bibinfo {author} {\bibfnamefont
  {A.}~\bibnamefont {Goldoni}}, \bibinfo {author} {\bibfnamefont
  {R.}~\bibnamefont {Larciprete}}, \ and\ \bibinfo {author} {\bibfnamefont
  {G.}~\bibnamefont {Paolucci}},\ }\href@noop {} {\bibfield  {journal}
  {\bibinfo  {journal} {Phys. Rev. B}\ }\textbf {\bibinfo {volume} {66}},\
  \bibinfo {pages} {075402} (\bibinfo {year} {2002})}\BibitemShut {NoStop}%
\bibitem [{nis()}]{nist-imfp}%
  \BibitemOpen
  \href@noop {} {}\bibinfo {note} {C. J. Powell and A. Jablonski, {\it NIST
  Electron Inelastic-Mean-Free-Path Database} - Version 1.2, National Institute
  of Standards and Technology, Gaithersburg, MD (2010).}\BibitemShut {Stop}%
\bibitem [{\citenamefont {de~Lima}\ \emph {et~al.}(2013)\citenamefont
  {de~Lima}, \citenamefont {de~Siervo}, \citenamefont {Landers}, \citenamefont
  {Viana}, \citenamefont {Goncalves}, \citenamefont {Lacerda},\ and\
  \citenamefont {Haberle}}]{deLima2013}%
  \BibitemOpen
  \bibfield  {author} {\bibinfo {author} {\bibfnamefont {L.~H.}\ \bibnamefont
  {de~Lima}}, \bibinfo {author} {\bibfnamefont {A.}~\bibnamefont {de~Siervo}},
  \bibinfo {author} {\bibfnamefont {R.}~\bibnamefont {Landers}}, \bibinfo
  {author} {\bibfnamefont {G.~A.}\ \bibnamefont {Viana}}, \bibinfo {author}
  {\bibfnamefont {A.~M.~B.}\ \bibnamefont {Goncalves}}, \bibinfo {author}
  {\bibfnamefont {R.~G.}\ \bibnamefont {Lacerda}}, \ and\ \bibinfo {author}
  {\bibfnamefont {P.}~\bibnamefont {Haberle}},\ }\href@noop {} {\bibfield
  {journal} {\bibinfo  {journal} {Phys. Rev. B}\ }\textbf {\bibinfo {volume}
  {87}},\ \bibinfo {pages} {081403(R)} (\bibinfo {year} {2013})}\BibitemShut
  {NoStop}%
\bibitem [{\citenamefont {de~Lima}\ \emph {et~al.}(2014)\citenamefont
  {de~Lima}, \citenamefont {Handschak}, \citenamefont {Schoenbohm},
  \citenamefont {Landers}, \citenamefont {Westphal},\ and\ \citenamefont
  {de~Siervo}}]{deLima2014_A}%
  \BibitemOpen
  \bibfield  {author} {\bibinfo {author} {\bibfnamefont {L.~H.}\ \bibnamefont
  {de~Lima}}, \bibinfo {author} {\bibfnamefont {D.}~\bibnamefont {Handschak}},
  \bibinfo {author} {\bibfnamefont {F.}~\bibnamefont {Schoenbohm}}, \bibinfo
  {author} {\bibfnamefont {R.}~\bibnamefont {Landers}}, \bibinfo {author}
  {\bibfnamefont {C.}~\bibnamefont {Westphal}}, \ and\ \bibinfo {author}
  {\bibfnamefont {A.}~\bibnamefont {de~Siervo}},\ }\href@noop {} {\bibfield
  {journal} {\bibinfo  {journal} {Chemical Communications}\ }\textbf {\bibinfo
  {volume} {50}},\ \bibinfo {pages} {13571} (\bibinfo {year}
  {2014})}\BibitemShut {NoStop}%
\end{thebibliography}

\end{document}